\documentclass{iaus}
\usepackage{graphicx}

\title[Mass-radius relationships of rocky exoplanets] 
{Mass-radius relationships of rocky exoplanets}

\author[F. Sohl, F.W. Wagner, H. Rauer]   
{Frank Sohl$^1$, Frank W. Wagner$^1$, \and Heike Rauer$^{1,2}$}

\affiliation{$^1$Institute of Planetary Research, German Aerospace Center (DLR), Berlin, Germany \\[\affilskip]
$^2$Center of Astronomy and Astrophysics, Technical University Berlin (TUB), Germany}

\pubyear{2012}
\volume{293}  
\pagerange{1--4}
\addtocounter{figure}{0}
\jname{Formation, Detection, and Characterization of Extrasolar Habitable Planets}
\editors{N. Haghighipour, ed.}
\begin{document}

\maketitle

\begin{abstract}
Mass and radius of planets transiting their host stars are provided by radial velocity and photometric observations. Structural models of solid exoplanet interiors are then constructed by using equations of state for the radial density distribution, which are compliant with the thermodynamics of the high-pressure limit. However, to some extent those structural models suffer from inherent degeneracy or non-uniqueness problems owing to a principal lack of knowledge of the internal differentiation state and/or the possible presence of an optically thick atmosphere. We here discuss the role of corresponding measurement errors, which adversely affect determinations of a planet's mean density and bulk chemical composition. Precise measurements of planet radii will become increasingly important as key observational constraints for radial density models of individual solid low-mass exoplanets or super-Earths.

\end{abstract}

\firstsection 
\section{Introduction}
The steadily growing number of transiting planet discoveries will allow us to characterize rocky exoplanets in terms of internal structure and atmospheric composition with important implications for their formation, orbital evolution, and possible habitability (see review by \cite[Haghighipour 2011]{haghighipour2011}). Numerical models of planetary interiors based on laboratory data of physical material properties are aimed at improving the general understanding of their origins, pathways of evolution, and diversity. In case of the terrestrial planets and satellites within the solar system, the resultant radial profiles of density and related material properties are required to satisfy geophysical observations and cosmochemical arguments for the compositions of major geochemical reservoirs such as the core, mantle, and crust (see \cite[Sohl \&\ Schubert 2007]{sohl2007}, \cite[Sohl 2010]{sohl2010}, and references therein). For rocky exoplanets, the numerical models have to be consistent with the observed planetary masses and radii measured from ground-based observations and space missions. Calculated models have been used to derive mass-radius relationships for exoplanets assuming a range of chemical compositions to gain insight into the bulk compositions and possible interior structures of these planets (e.g., \cite[Valencia 2011]{valencia2011}, and references therein). \cite{wagner2011a} have reinvestigated mass-radius relationships for rocky exoplanets using equations of state that are compliant with the thermodynamics of the high-pressure limit of a given material.

The mean density is the main indicator of the bulk composition of solid planets. The purpose of this paper is to discuss the role of mass and radius measurement errors for determinations of a planet's mean density and bulk chemical composition by using calculated relationships between radius and mass of solid low-mass exoplanets.

\section{Method}
Since there are usually fewer constraints than unknowns, even basic interior structure models that would involve only two or three chemically homogeneous layers of constant density suffer from inherent degeneracy. In addition, structural models of low-mass exoplanets have non-uniqueness problems because of their unknown differentiation state, and/or the possible presence of an optically thick atmosphere (\cite[Valencia 2011]{valencia2011}), and the extrapolation of equations of state of mineral phase assemblages to extremely high pressures (\cite[Swift \etal\ 2012]{swift2012}). 
Further uncertainties are related to pressure-induced mantle phase transitions, the stability field of post-perovskite, the possible presence of additional high-pressure silicate phases, and the physical state of core iron alloys (\cite[Fortney \etal\ 2009]{fortney2009}). This has important implications for mass-radius relationships and their usage for the characterization of low-mass exoplanets in terms of bulk composition and interior structure and the possible existence of self-sustained magnetic fields (\cite[Wagner \etal\ 2011b]{wagner2011b}).

To obtain relationships between planetary radius $R_p$ and total mass $M_p$ for solid planets according to 
\begin{equation}
R_p = c M_p^{\beta} ,
\label{eq:mr}
\end{equation}
where $\beta$ is the scaling exponent, we first solve the structural equations for mass and momentum assuming hydrostatic equilibrium (e.g., \cite[Wagner \etal\ 2012]{wagner2012})
\begin{equation}
\label{eqs:basicdiffs}
\frac{d m}{d r}  =   4 \pi r^2 \rho(r); \hspace{1cm}
\frac{d P}{d r}  =   - \frac{G m}{r^2}\rho(r) ,
\end{equation}
where $G$ is the gravitational constant, together with an equation of state (EoS) for the local density $\rho$ according to
\begin{eqnarray} \label{eqs:eos}
 \rho(r) & = & f_{EoS}(P).
 \label{eq:eos}
\end{eqnarray}
Here, we adopt the generalized Rydberg EoS with material parameters that are valid for iron (Fe), silicate (MgSiO$_{3}$), and water ice (H$_{2}$O). This EoS has the advantage of being compliant with the high-pressure limit of a given material, thereby describing the radial distribution of density $\rho(r)$ up to extremely large compression ratios of solid planetary materials (see \cite[Wagner \etal\ 2011a]{wagner2011a}, and references therein). We further neglect the thermal pressure contribution because of its minor importance for massive exoplanet interiors (e.g., \cite[Seager \etal\ 2007]{seager2007}). Successful solutions of the structural equations are required to satisfy the boundary conditions 
\begin{equation} \label{eqs:bounds}
 m(0)  =  0, \hspace{0.2in}
 P(R_{p})  =   0, \hspace{0.2in}
 P(r_{pt})  =   f_{pt}
\end{equation}
at the planet's center ($r=0$), surface ($r=R_p$), and pressure-induced phase transition boundaries ($r=r_{pt}$). We are taking into account transition pressures $f_{pt}$ for the $\alpha$- and $\epsilon$-phase of iron, the perovskite and post-perovskite phase of MgSiO$_{3}$, and the low- and high-pressure phases of water ice.

\begin{table}[ht]
\begin{center}
\caption{Scaling exponent $\beta$ for solid exoplanets in the mass range of $1$~to~$10$~M$_\oplus$.}
\label{tab:scale}
{\scriptsize
\begin{tabular}{lcccc}\hline
{\bf Composition}   & \cite{valencia2006}  & \cite{sotin2007}  & \cite{wagner2011a} \\[2pt]\hline
Earth-like          &   0.262              &     0.274         &   0.267           \\
Ocean planet        &   0.244              &     0.275         &   0.261           \\
Mercury-type        &    none              &      none         &   0.269           \\[5pt]\hline
\end{tabular}
}
\end{center}
\vspace{1mm}
\scriptsize{ 
{\it Note:} Assumed is a radius-mass relationship according to $R_p = c M_p^{\beta}$ with scaling exponent $\beta$.}
\\[2pt]
\end{table}

Second, a power law fit is performed to adjust the scaling exponent $\beta$ in eq.~\ref{eq:mr} for solid low-mass exoplanets in the mass range of $1$~to~$10$~M$_\oplus$. Tab.~\ref{tab:scale} provides a comparison of published scaling law exponents for solid low-mass exoplanets of various compositions. Because of the large compression ratios involved, those are substantially less than $\beta=1/3$ in case of a homogeneous density distribution.

\begin{figure}[tt]
\begin{center}
  \includegraphics[width=0.75\textwidth]{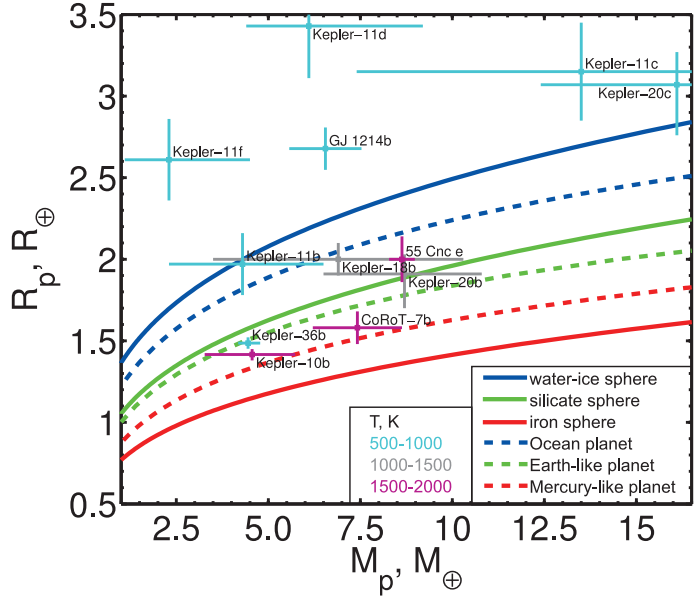}\\
  \caption{\small Mass-radius diagram for planets with different bulk compositions compared to currently known low-mass exoplanets in Earth units. We divide equilibrium surface temperatures into three domains from 500 to 1000 K; 1000 to 1500 K; and 1500 to 2000 K. While the solid curves denote homogeneous, self-compressible solid spheres of water ice, silicate rock, and iron, respectively, the dashed curves exhibit differentiated models of intermediate bulk compositions.}
  \label{mr10}
\end{center}
\end{figure}

The mean density of a spherical planet is given by 
\begin{equation}
\bar{\rho} = \frac{3}{4\pi} \frac{M_p}{R_p^3} ,
\label{eq:rho}
\end{equation}
where, in general, mass $M_p$ and radius $R_p$ of planets transiting their host stars are provided independently from each other by radial velocity and photometric observations.
We therefore employ an error propagation analysis according to 
\begin{equation}
{\Delta \bar{\rho}} = \left[ \left(\frac{\partial\bar{\rho}}{\partial M_p}\right)^2 {\Delta M_p}^2 + \left(\frac{\partial\bar{\rho}}{\partial R_p}\right)^2 {\Delta R_p}^2 \right]^{\frac{1}{2}}
\label{eq:err0}
\end{equation}
and obtain, upon substitution of the radius-mass relationship given in eq.~\ref{eq:mr}, the propagated relative error in mean density
\begin{equation}
\frac{\Delta \bar{\rho}}{\bar{\rho}} = \left[ \frac{1 + 9\beta^2}{2} \left(\frac{\Delta M_p}{M_p}\right)^2 + \frac{9 + \beta^{-2}}{2} \left(\frac{\Delta R_p}{R_p}\right)^2 \right]^{\frac{1}{2}}
\label{eq:err1}
\end{equation}
as a function of the scaling law exponent $\beta$ and the observational uncertainties of the mass and radius determinations. The latter can be expressed in terms of some key observables, namely
\begin{equation}
\label{eq:obs}
 \frac{\Delta K^{\ast}}{K^{\ast}} = \frac{\Delta M_p}{M_p} ; \hspace{1cm}
 \frac{\Delta \delta^{\ast}}{\delta^{\ast}} = 2 \frac{\Delta R_p}{R_p} ,
\end{equation}
where $K^{\ast}$ and $\delta^{\ast}$ denote radial velocity semi-amplitude and transit depth of the host star, respectively.
Upon insertion of $\beta=1/3$ in eq.~\ref{eq:err1} for a homogeneous density distribution, the relative error in mean density can be written as   
\begin{equation}
\frac{\Delta \bar{\rho}}{\bar{\rho}} = \left[ \left(\frac{\Delta M_p}{M_p}\right)^2 + 9\left(\frac{\Delta R_p}{R_p}\right)^2 \right]^{\frac{1}{2}} ,
\label{eq:err2}
\end{equation}
which is a valid approximation for small terrestrial planets and moons ($M_p \ll 1~{\rm M}_\oplus$) that are subject to low internal pressures and negligible compression ratios.

\begin{figure}[tt]
\begin{center}
  \includegraphics[width=0.8\textwidth]{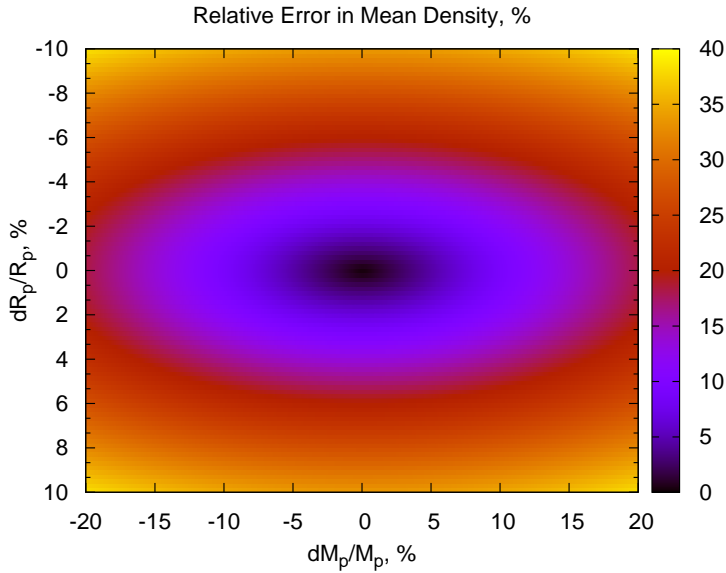}\\
  \caption{\small Relative error in mean density, ${\Delta \bar{\rho}}/{\bar{\rho}}$, as a function of relative uncertainties in planetary mass, ${\Delta M_p}/M_p$, and radius, ${\Delta R_p}/R_p$, as given by eq.~\ref{eq:err1}. Assumed is a radius-mass relationship according to $R_p = c M_p^{\beta}$ with scaling exponent $\beta = 0.27$, representative for solid low-mass exoplanets or super-Earths.}
  \label{error}
\end{center}
\end{figure}

\section{Results and Discussion}
The mass-radius diagram shown in Fig.~\ref{mr10} exhibits the remarkable compositional diversity of known transiting low-mass exoplanets. Whereas the low density of GJ~1214b, that has no analogue in the solar system, is attributed to the presence of an extended optically thick hydrogen atmosphere (\cite[Rogers \&\ Seager 2011]{rogers2011}), it has been suggested that carbon-rich solids may predominate the bulk composition of 55 Cancri~e (\cite[Madhusudhan \etal\ 2012]{madhusudhan2012}). CoRoT-7b and Kepler-10b, on the other hand, are thought to be composed of silicate rock and iron with the latter concentrated in massive metallic cores (\cite[Wagner \etal\ 2012, Swift \etal\ 2012]{wagner2012,swift2012}). These compositional trends are roughly followed by the calculated equilibrium surface temperatures, with relatively hot planetary environments along and below the silicate line and colder but still hot environments along and above the water ice line. Furthermore, it is important to note that solid planets with masses above $10$~M$_\oplus$ were not discovered to date, although transit detections should be readily feasible owing to the relatively large planet radius.

\begin{figure}[tt]
\begin{center}
  \includegraphics[width=0.95\textwidth]{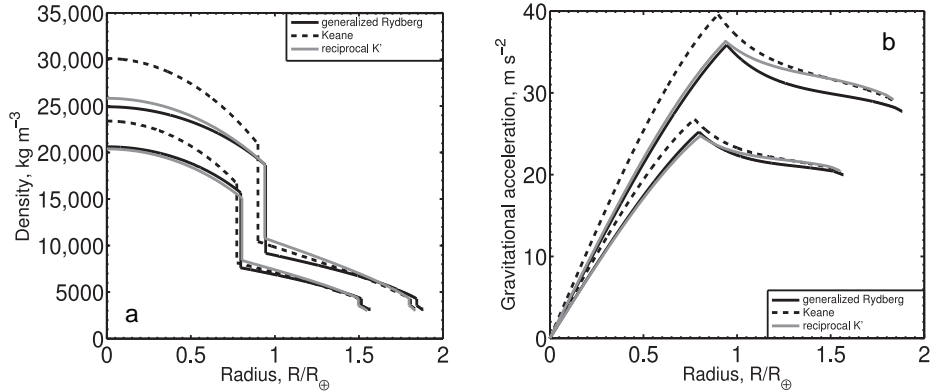}\\
  \caption{\small Radial distributions of (a) density and (b) gravitational acceleration for solid model planets of $5$~M$_{\oplus}$ (lower set of model curves) and $10$~M$_{\oplus}$ (upper set of model curves), respectively. Planet bulk composition is held fixed to that of the Earth. Using the generalized Rydberg, Keane, and reciprocal $K'$ EoS, the resultant surface radii differ by less than 2\% from each other (adopted from \cite[Wagner \etal\ 2011a]{wagner2011a}).}
  \label{eos}
\end{center}
\end{figure}

Inspection of Tab.~\ref{tab:scale} suggests that the scaling exponent $\beta$ is similar for radius-mass relationships of solid exoplanets in the mass range of $1$~to~$10$~M$_\oplus$, irrespective of their bulk compositions. To calculate ellipsoidal distributions of the relative error in mean density, ${\Delta \bar{\rho}}/{\bar{\rho}}$, we thereby assume an average scaling exponent $\beta=0.27$, being representative for solid low-mass exoplanets or super-Earths. From inspection of Fig.~\ref{error}, it is seen that the accurate determination of mean planetary density depends on the observational uncertainties of both planetary mass and radius. The performance comparison of ongoing and proposed transit missions given in Tab.~\ref{tab:mission} hints at the expected accuracy of corresponding planet density determinations. Owing to the variable slope of mass-radius relations, planetary mean density in the giant-planet mass range is mainly constrained by precise radius determinations, whereas mass and radius pose equally important constraints on structural models of low-mass planets.

\begin{table}[ht]
\begin{center}
\caption{Performance of ongoing and proposed transit missions.}
\label{tab:mission}
{\scriptsize
\begin{tabular}{lccccc}\hline
{\bf Uncertainty}                 & CoRoT$^{a}$ & Kepler$^{b}$ & CHEOPS$^{c}$     & PLATO$^{d}$ \\[2pt]\hline
radius, $\frac{\Delta R_p}{R_p}$  & $\pm 0.06$  & $\pm 0.02$   & $\pm (0.02-0.1)$ & $\pm 0.02$  \\
mass, $\frac{\Delta M_p}{M_p}$    & $\pm 0.15$  & $\pm 0.27$   & $\pm 0.10$       & $\pm 0.10$  \\
                                  & (CoRoT-7b)  & (Kepler-10b) &                  &             \\[5pt]\hline
\end{tabular}
}
\end{center}
\vspace{1mm}
\scriptsize{
{\it References:}\\
$^{a)}$ \protect\cite{bruntt2010};
$^{b)}$ \protect\cite{batalha2011};
$^{c)}$ CHEOPS proposal ({\tt http://cheops.unibe.ch});
$^{d)}$ PLATO M3 proposal ({\tt http://www.oact.inaf.it/plato/PPLC/Home.html}). 
}
\\[2pt]
\end{table}

In Fig.~\ref{eos}, structural model determinations of atmosphere-less solid planets are compared for different equations of state and fixed Earth-like bulk compositions. It is seen that the trade-off in calculated planet radius is smaller than present-day measurement uncertainties from transit photometry, indicating that mass-radius relationships of solid exoplanets are sufficiently robust to infer their bulk compositions (\cite[Wagner \etal\ 2011a]{wagner2011a}).
Since mass determinations using the radial velocity method are currently limited to an accuracy of about $\pm 10$ percent, precise measurements of planet radius will become increasingly important as key observational constraint for radial density models of individual low-mass exoplanets. 
At any rate, the precise and reliable characterization of planet host stars in terms of stellar radius, mass, and age is prerequisite to infer the mean density and thereby composition of low-mass exoplanets or super-Earths with confidence. This can be accomplished by using asteroseismology on sufficiently bright and small target stars (e.g., \cite[Rauer \etal\ 2011]{rauer2011}).

\section{Conclusions}
Current detection limits of ground-based observational methods have limited the discovery of solid exoplanets to only a few, although low-mass planets beyond the solar system should be quite abundant (e.g., \cite[Marcy \etal\ 2011]{marcy2011}). Mass-radius relationships based on numerical models of solid exoplanet interiors are sufficiently robust to infer a planet's bulk composition from accurate determinations of its mean density. The future detection of transiting mini-Neptunes and super-Earths will provide fundamental information to better constrain their bulk compositions and possible interior structures in terms of metallic cores, silicate/carbon-rich mantles, water-ice/liquid shells, and/or atmosphere mass fractions.

\begin{acknowledgment}
\noindent
This research has been supported by the Helmholtz Alliance "Planetary Evolution and Life".
\end{acknowledgment}

\end{document}